\documentclass[preprint,secnumarabic,amssymb,amsmath, nobibnotes, aps, prd]{revtex4-1}
\usepackage{bbm} 
\usepackage{graphicx}
\usepackage{hyperref}
\usepackage{bm}

\begin{document}
\title{\boldmath Precision calculation of the axion-nucleon coupling in chiral perturbation theory}

\author{Thomas Vonk}
 \email{vonk@hiskp.uni-bonn.de}
 \affiliation{Helmholtz-Institut f\"{u}r Strahlen- und Kernphysik and Bethe Center for Theoretical Physics,
   Universit\"{a}t Bonn, D-53115 Bonn, Germany}
 
 \author{Feng-Kun Guo}
 \email{fkguo@itp.ac.cn}
 \affiliation{CAS Key Laboratory of Theoretical Physics, Institute of Theoretical Physics, Chinese Academy
   of Sciences, Beijing 100190, China}
 \affiliation{School of Physical Sciences, University of Chinese Academy of Sciences,
    Beijing 100049, China}

\author{Ulf-G. Mei{\ss}ner}
 \email{meissner@hiskp.uni-bonn.de}
 \affiliation{Helmholtz-Institut f\"{u}r Strahlen- und Kernphysik and Bethe Center for Theoretical Physics,
   Universit\"{a}t Bonn, D-53115 Bonn, Germany}
 \affiliation{Institute for    Advanced Simulation, Institut f\"ur Kernphysik and J\"ulich Center for Hadron
   Physics,  Forschungszentrum J\"ulich, D-52425 J\"ulich, Germany}
 \affiliation{Tbilisi State University, 0186 Tbilisi, Georgia}

\date{\today}
\begin{abstract}We derive the axion-nucleon interaction Lagrangian in heavy baryon
  chiral perturbation theory up to next-to-next-to-leading order. The effective axion-nucleon
  coupling is calculated to a few percent accuracy.
\end{abstract}

\maketitle

\section{Introduction}

More than forty years after the proposal to add another symmetry in QCD, viz. the Peccei-Quinn (PQ) symmetry
U(1)$_\text{PQ}$ \cite{Peccei:1977hh,Peccei:1977ur}, the pseudo-Nambu-Goldstone boson resulting from the
spontaneous breakdown of this symmetry, the QCD axion, remains one of the most favored candidates for a
Beyond the Standard Model (BSM) particle. The reasons are manifold: Originally introduced as a resolution of the
so-called strong-$CP$ problem, i.e. the question why the observable QCD vacuum angle $\bar{\theta}=
\theta_\text{QCD}+\operatorname{arg}\operatorname{det} \mathcal{M}_q$ (with $\mathcal{M}_q$ the quark
mass matrix) is such a small quantity (current  measurements of the neutron electric dipole moment
imply $|\bar{\theta}|\lesssim 10^{-11}$ \cite{baluni,GuoEtAlNEDM,Dragos:2019oxn}), its experimental
detection would not only unequivocally solve the strong-$CP$ problem, but potentially also provide
an answer (or complement the answer) on the question of the nature of the cosmological dark matter,
another pressing issue in contemporary physics research. At the same time a model with
PQ symmetry breaking can lead to massive Majorana or Dirac neutrinos depending on the choice of assigning
PQ charges to the SM particles and Higgses \cite{Mohapatra:1982tc,Langacker:1986rj,Shin:1987xc,He:1988dm,Chen:2012baa,Bertolini:2014aia,Gu:2016hxh,Suematsu:2017kcu,Reig:2018yfd,Peinado:2019mrn}. Moreover, the fact that axions
and PQ symmetries arise quite naturally in superstring theory \cite{Svrcek:2006yi} increases their popularity further.

If the QCD axion indeed exists, its couplings to Standard Model particles, i.e. matter particles and gauge bosons, and hence to composite particles as nucleons, must be very weak, because these are controlled by the very large axion decay constant $f_a$. Currently only lower and upper bounds on these couplings can be given.
If these bounds are determined from nuclear processes, the exactness of the determination of $f_a$ then strongly
depends on the accuracy of our knowledge on the effective axion-nucleon coupling strength.

The leading order axion-nucleon coupling has been derived long ago in Ref.~\cite{Donnelly:1978ty} for the
Peccei-Quinn-Weinberg-Wilczek (PQWW) axion \cite{Peccei:1977hh,Peccei:1977ur,Weinberg:1977ma,Wilczek:1977pj}
based on current algebra techniques, and --- building upon the same work --- in
Refs.~\cite{Kaplan:1985dv,Srednicki:1985xd,Georgi:1986df,Chang:1993gm} in a more general manner. Here, we strive
for deriving the axion-nucleon interaction in heavy baryon chiral perturbation theory (HBCHPT) for an arbitrary
axion model coupling to hadrons as well as, in particular, for the Kim-Shifman-Vainstein-Zakharov (KSVZ) axion
\cite{Kim:1979if,Shifman:1979if} and the Dine-Fischler-Srednicki-Zhitnitskii (DFSZ) axion
\cite{Dine:1981rt,Zhitnitskii:1980}. To leading order, this has been done in Ref.~\cite{diCortona:2015ldu}, so we
extend their analysis to sub-leading orders, because more precise estimations of the axion-nucleon coupling
allow for improved determinations of astrophysical constraints on the axion mass, or, equivalently, the axion
decay constant, e.g. from the axion bremsstrahlung processes \cite{Iwamoto:1984ir,Mayle:1987as,Brinkmann:1988vi,Raffelt:1987yt,Keil:1996ju,Hanhart:2000ae,Chang:2018rso,Carenza:2019pxu}.

Our work is organized as follows: In Sec.~\ref{sec:aq}, we recapitulate the 
interaction Lagrangian between quarks and the axion. Then, in Sec.~\ref{sec:aN}, we derive the axion-nucleon
interaction to the third order, that is including all terms up to next-to-next-to-leading order and pion loop
contributions. We also give the numerical values of the
axion coupling to neutrons and protons. We end with a short summary in Sec.~\ref{sec:summ}.

\section{Axion-quark interaction Lagrangian}
\label{sec:aq}

Consider the QCD Lagrangian including the axion field $a(x)$ at energies below the PQ
scale \cite{Peccei:1977hh,Peccei:1977ur}  with $q=(u,d,s,c,b,t)^\mathrm{T}$,  
\begin{equation}
  \mathcal{L}_\mathrm{QCD} =  \mathcal{L}_\mathrm{QCD, 0} - \bar{q}\mathcal{M}_q q 
  + \frac{a}{f_a} \frac{g^2}{16\pi^2}\operatorname{Tr}\left[G_{\mu\nu}\tilde{G}^{\mu\nu}\right]
  + \frac{\partial^\mu a}{2f_a} J_\mu^\mathrm{PQ}\ ,
  \label{eq:startinglagrangian}
\end{equation}
where $\mathcal{L}_\mathrm{QCD, 0}$ contains all terms that are not of interest in what follows, including
the axion-photon interaction term \cite{Kim:1986ax,Kim:2008hd,diCortona:2015ldu}. Furthermore, $\mathcal{M}_q
= \operatorname{diag}\left(m_u,m_d,m_s,m_c,m_b,m_t\right)$ is the quark mass matrix, $f_a$ is the axion decay
constant, $g$ the strong interaction coupling constant, $G_{\mu\nu}=G^a_{\mu\nu}\lambda^a/2$ is the conventional gluon field strength tensor with $\lambda^a$ the Gell-Mann matrices,
and $\tilde{G}_{\mu\nu}=\frac{1}{2}\epsilon_{\mu\nu\alpha\beta}G^{\alpha\beta}$ its dual, where the trace hence acts in the color
space. The PQ current is given by
\begin{equation}
J_\mu^\mathrm{PQ} = f_a \partial_\mu a + \bar{q}\gamma_\mu\gamma_5 \mathcal{X}_q q \ ,
\end{equation}
from which the first term gives rise to the kinetic term of the axion, whereas the second term describes the
axion-quark interactions proportional to the model-dependent coupling constants combined in the matrix
$\mathcal{X}_q=\operatorname{diag}(X_q)$ acting in the flavor space. These are given by
\begin{align}
\label{eq:couplignconstantsmodeldepending}
\begin{split}
X_q^\mathrm{KSVZ} & = 0 \ , \\
X_{u,c,t}^\mathrm{DFSZ} & = \dfrac{1}{3} \dfrac{x^{-1}}{x+x^{-1}} = \dfrac{1}{3}\sin^2\beta\ , \\
X_{d,s,b}^\mathrm{DFSZ} &=\dfrac{1}{3} \dfrac{x}{x+x^{-1}}=\dfrac{1}{3}\cos^2\beta = \dfrac{1}{3}
- X_{u,c,t}^\mathrm{DFSZ} \ , 
\end{split}
\end{align}
for the KSVZ axion and the DFSZ axion, respectively, and $x=\cot\beta$ is the ratio of the vacuum expectation
values (VEVs) of the two Higgs doublets. We exclude the PQWW axion from the analysis since it has been ruled
out experimentally \cite{Donnelly:1978ty,Kim:2008hd}.

Note that we do not integrate out the heavy quarks from the beginning. As, depending on the model, the axion
couplings to heavy quarks are quite a possibility, they might contribute to the axion-nucleon interactions due to sea
quark effects. Additionally, as the couplings $X_q$ are scale-dependent quantities \cite{diCortona:2015ldu}, running
effects enter the couplings to nucleons, which can only be recovered if the axion interactions with heavy quarks
are taken along within the calculations, to wit in the form of isoscalar currents.

It is advisable to perform an axial rotation on the quark fields in order to remove the term
$\propto a\operatorname{Tr}\left[G_{\mu\nu}\tilde{G}^{\mu\nu}\right]$ in Eq.~\eqref{eq:startinglagrangian}
by transforming
\begin{equation}
q \to \exp\left({i\gamma_5\frac{a}{2f_a}\mathcal{Q}_a}\right)\,q 
\end{equation} 
with
\begin{equation}
  \mathcal{Q}_a=\frac{\mathcal{M}_q^{-1}}{\operatorname{Tr}\mathcal{M}_q^{-1}}\approx
  \frac{1}{1+z+w}\operatorname{diag}\left(1,z,w,0,0,0\right)\ ,
\end{equation} 
where $z=m_u/m_d$ and $w=m_u/m_s$. The particular form of $\mathcal{Q}_a$ has been chosen in order to avoid the
$\pi^0$-$a$ mass mixing~\cite{Georgi:1986df}. 

With that transformation, the Lagrangian can be written as
\begin{equation}
\label{eq:shiftedlagrangian}
\mathcal{L}_\mathrm{QCD}^\prime = \mathcal{L}_\mathrm{QCD, 0}^\prime -  \left(\bar{q}_L\mathcal{M}_a q_R
+ \text{h.c.} \right) + \frac{\partial^\mu a}{2f_a} J_\mu^a\ ,
\end{equation}
where now the non-derivative axion-quark interactions are entirely shifted into the phase of the mass matrix,
\begin{equation}\label{eq:Mass:a}
\mathcal{M}_a=\exp\left({i\frac{a}{f_a}\mathcal{Q}_a}\right) \, \mathcal{M}_q\ ,
\end{equation}
whereas the derivative axion-quark interactions are present in the coupling to the axion current
\begin{equation}\label{eq:currentaftertrafo}
J_\mu^a = J_\mu^\mathrm{PQ} - \bar{q}\gamma_\mu\gamma_5\mathcal{Q}_a q\ ,
\end{equation}
which is now anomaly-free \cite{Bardeen:1977bd,Donnelly:1978ty}. The terms in $J_\mu^a$ must be split into
isoscalar and isovector pieces in order to translate it later into an effective field theory (EFT) language.
Consider the two-dimensional subspace of Eq.~\eqref{eq:currentaftertrafo} with $q=(u,d)^\mathrm{T}$: 
\begin{align}
J_\mu^{a,ud} & = f_a \partial_\mu a + \bar{q}\gamma_\mu\gamma_5 \left(\mathcal{X}_q-\mathcal{Q}_a \right) q \nonumber\\
& = f_a \partial_\mu a  + c_{u-d} \bar{q}\gamma_\mu\gamma_5 \tau_3 q + c_{u+d} \bar{q}\gamma_\mu\gamma_5 q\ ,
\label{eq:currentsplitted}
\end{align}
where $\tau_3$ is the conventional third Pauli matrix and we have introduced the abbreviations
\begin{align}
\label{eq:X3andXs}
\begin{split}
c_{u-d} &= \frac{1}{2}\left(X_u-X_d-\frac{1-z}{1+z+w}\right)  ,  \\
c_{u+d} &= \frac{1}{2}\left(X_u+X_d-\frac{1+z}{1+z+w}\right)  .
\end{split}
\end{align}
Setting furthermore
\begin{equation}\label{eq:cqs}
c_s=X_s-\frac{w}{1+z+w}\ , \qquad c_{c,b,t}=X_{c,b,t}
\end{equation}
and inserting Eq.~\eqref{eq:currentsplitted} into Eq.~\eqref{eq:shiftedlagrangian}, one finds the general
axion-quark interaction Lagrangian
\begin{align}
  \mathcal{L}_{a\text{--}q} = & - \bar{q}_L\mathcal{M}_a q_R + \text{h.c.}\nonumber \\ &
  + \left( \bar{q}\gamma^\mu \frac{\partial_\mu a}{2f_a} \left(c_{u-d}\tau_3+ c_{u+d} \bm{1}\right)
  \gamma_5 q\right)_{q=(u,d)^\mathrm{T}} 
  + \left(  c_q \bar{q}\gamma^\mu \frac{\partial_\mu a}{2f_a} \gamma_5 q\right)_{q=(s,c,b,t)^\mathrm{T}}  \ ,
  \label{eq:a-q-lagrangian}
\end{align}
which is now expressed in a suitable basis so that the isovector and isoscalar parts of the axion-nucleon
interaction can easily be extracted.

\section{Axion-nucleon interaction in heavy baryon chiral perturbation theory}
\label{sec:aN}
\subsection{The Lagrangian}

We construct the HBCHPT Lagrangian with the additional axion field and its interactions by adapting the one
developed in Ref.~\cite{Fettes:1998ud} (including the notation) and adding additional terms allowed from
symmetries containing the isoscalar axial currents. Usually, the axial currents entering the HBCHPT
Lagrangian as external sources are taken to be traceless in order to avoid subtleties arising from the
U(1)$_A$ anomaly. However, here our model is anomaly-free by construction and we now have to add the
isoscalar axial currents appearing separately in the Lagrangian in Eq.~\eqref{eq:a-q-lagrangian} that are not
traceless. This is done in complete analogy to the traceless axial currents.

We introduce
\begin{equation}
u=\sqrt{U}=\exp\left(i\frac{\pi^a\tau_a}{2F_\pi}\right)
\end{equation}
which contains the three pseudo-Nambu-Goldstone bosons of the spontaneously broken chiral symmetry,
with the index $a=1,2,3$ and summation implied. Furthermore, $F_\pi$ is the pion decay constant in the chiral limit,
for which we will take the physical value 92.1\,MeV for the difference to the chiral limit value only amounts
to effects of higher orders than those considered here. The isovector axial current $a_\mu$ enters the theory
by means of the chiral connection $\Gamma_\mu$ of the covariant derivative,
\begin{equation}
  D_\mu= \partial_\mu + \Gamma_\mu= \partial_\mu + \frac{1}{2} \left[ u^\dagger \partial_\mu u
    + u \partial_\mu u^\dagger -i u^\dagger a_\mu u  + i u a_\mu u^\dagger \right] , 
\end{equation}
the so-called vielbein,
\begin{equation}
  u_\mu = i\left[ u^\dagger \partial_\mu u - u \partial_\mu u^\dagger - i u^\dagger
  a_\mu u - i u a_\mu u^\dagger \right] ,
\end{equation}
and the field-strength tensor,
\begin{equation}
F_{\mu\nu}^\text{L,R} = \mp \partial_\mu a_\nu \pm \partial_\nu a_\mu - i \left[a_\mu, a_\nu\right] ,
\end{equation}
where we have set $v_\mu=0$ for the external vector field. Note that the field-strength tensor vanishes in
the present model because $a_\mu \propto \partial_\mu a\, \tau_3$ as can be read off from the
Lagrangian~\eqref{eq:a-q-lagrangian}. Introducing thus the isoscalar axial current $a_{\mu,i}^s\propto
c_i \partial_\mu a\, \mathbbm{1}$, we can construct similar objects: a connection
\begin{equation}
\tilde{\Gamma}_\mu = \frac{1}{2} \left[-i u^\dagger a_{\mu,i}^s u  + i u a_{\mu,i}^s u^\dagger \right]=0 , 
\end{equation}
which vanishes due to $a_{\mu,i}^s u=u a_{\mu,i}^s$, and a vielbein equivalent
\begin{equation}
  \tilde{u}_{\mu,i} = i\left[ - i u^\dagger
  a_{\mu,i}^s u - i u a_{\mu,i}^s u^\dagger \right] = 2 a_{\mu,i}^s ~.
\end{equation}
The corresponding field strength tensors, of course, vanish as in the case of the isovector axial current.
The index  $i=(u+d,s,c,b,t)$ runs over all isoscalar quark combinations, cf. Eq.~\eqref{eq:a-q-lagrangian}.
Furthermore, we need as the last building block, 
\begin{equation}
\chi_\pm= u^\dagger\chi u^\dagger \pm u \chi^\dagger u\ ,
\end{equation}
where $\chi=2B(s-ip)$ includes the external scalar and pseudoscalar fields $s(x)$ and $p(x)$, and
$B$ is a constant related to quark condensate $\Sigma=-\langle\bar{u}u\rangle$ via
$B=\lim_{m_u,m_d\to 0}(\Sigma/F^2)$. Collecting the proton and neutron fields in the isodoublet
$N (x) =(p,n)^\mathrm{T}$, the most general HBCHPT Lagrangian up to order $\mathcal{O}\left(p^3\right)$
in the low-energy expansion,
\begin{equation}
  \mathcal{L}_{\pi N}^{}  = \mathcal{L}_{\pi N}^\mathrm{(1)}  + \mathcal{L}_{\pi N}^\mathrm{(2)}
  + \mathcal{L}_{\pi N}^\mathrm{(3)}  + \dots,
\end{equation}  
reads 
\begin{align}
  \mathcal{L}_{\pi N} = \bar{N} \Biggl\{ & 
  		iv\cdot D+g_A S\cdot u + g_0^i S\cdot \tilde{u}_i 	-\frac{ig_A}{2m}\left\{ S\cdot D,v\cdot u \right\}-\frac{ig_0^i}{2m}\left\{ S\cdot D,v\cdot \tilde{u}_i \right\}\nonumber\\ &
  		+\frac{g_A}{8m^2} \left[D^\mu,\left[D_\mu, S\cdot u\right]\right]+\frac{g_0^i}{8m^2} \left[D^\mu,\left[D_\mu, S\cdot \tilde{u}_i\right]\right]\nonumber\\ &
  		-\frac{g_A}{4m^2} v\cdot\overset{\leftarrow}{D}\, S\cdot u\, v\cdot D  -\frac{g_0^i}{4m^2} v\cdot\overset{\leftarrow}{D}\, S\cdot \tilde{u}_i\, v\cdot D\nonumber\\ &
  		-\frac{g_A}{4m^2}\left(\left\{S\cdot D, v\cdot u\right\}v\cdot D + \text{h.c.}\right)-\frac{g_0^i}{4m^2}\left(\left\{S\cdot D, v\cdot \tilde{u}_i\right\}v\cdot D + \text{h.c.}\right)\nonumber\\ &
  		-\frac{g_A}{8m^2}\left(S\cdot u\, D^2 + \text{h.c.}\right)-\frac{g_0^i}{8m^2}\left(S\cdot \tilde{u}_i\, D^2 + \text{h.c.}\right)\label{eq:LagrangianNPi}\\ &
  		-\frac{g_A}{4m^2}\left(S\cdot\overset{\leftarrow}{D}\, u\cdot D +\text{h.c.}\right)-\frac{g_0^i}{4m^2}\left(S\cdot\overset{\leftarrow}{D}\, \tilde{u}_i\cdot D +\text{h.c.}\right)\nonumber\\ &
  		+d_{16}(\lambda) S\cdot u \operatorname{Tr}\left[\chi_+\right]+d_{16}^i(\lambda) S\cdot \tilde{u}_i \operatorname{Tr}\left[\chi_+\right]\nonumber+d_{17} S^\mu \operatorname{Tr}\left[u_\mu\chi_+\right]\nonumber\\ &
  		+id_{18} S^\mu\left[D_\mu,\chi_{-}\right] + id_{19}S^\mu\left[D_\mu,\operatorname{Tr}[\chi_{-}]\right]\nonumber\\ & 
  		+\tilde{d}_{25}(\lambda)v\cdot\overset{\leftarrow}{D}\, S\cdot u\, v\cdot D+\tilde{d}^i_{25}(\lambda)v\cdot\overset{\leftarrow}{D}\, S\cdot \tilde{u}_i\, v\cdot D\nonumber\\
  & +\tilde{d}_{29}(\lambda)\left(S^\mu\left[v\cdot D, u_\mu\right] v\cdot D + \text{h.c.}\right)+\tilde{d}_{29}^i(\lambda)\left(S^\mu\left[v\cdot D, \tilde{u}_{\mu,i}\right] v\cdot D + \text{h.c.}\right) \biggr\} N\ ,\nonumber
\end{align}
where we only show terms that finally lead to interaction vertices with only one single axion, because
interactions containing $n$ axions are suppressed by factors $1/f_a^n$ and can hence be neglected. Note that
there is no isoscalar counterpart to the $d_{17}$ term proportional to a low-energy constant (LEC) $d_{17}^i$, since such a term would
have the same structure as the $d_{16}^i$ term, because $\tilde{u}_i\propto \mathbbm{1}$, and thus not independent. 
The terms in the first line are the leading order and next-to-leading order terms,
while all the other terms are the next-to-next-to-leading order terms. In the Lagrangian~\eqref{eq:LagrangianNPi},
$v_\mu$ is the nucleon four-velocity and $N=N_v$ are  velocity-dependent nucleon fields with mass $m$.
Strictly speaking, $m$ is the nucleon mass in the two-flavor chiral limit, often denoted as $\mathring{m}_N$. We
will suppress the index $v$ in what follows.  The axial couplings $g_A$ and $g_0^i$ should also be taken in
the chiral limit, but we will later match them with the nucleon matrix elements $\Delta q$, which refer
to the physical values of the quark masses, see below. $S_\mu$ is the covariant spin-operator, 
\begin{equation}
S_\mu = \frac{i}{2}\gamma_5\sigma_{\mu\nu} v^\nu\ ,
\end{equation}
which has the following properties in $d$ dimensions needed later, employing  dimensional regularization
to deal with the appearing divergences:
\begin{equation}
S\cdot v  = 0 ~, \quad
S^2  = \frac{1-d}{4}~, \quad
\left\{S_\mu,S_\nu\right\}  = \frac{1}{2}\left(v_\mu v_\nu - g_{\mu\nu}\right) . \label{eq:Spinproperties}
\end{equation}
Besides the parameters already mentioned, a number of new LECs $d^{(i)}_n$ and
$\tilde{d}^{(i)}_n$ appear, of which some depend on the scale $\lambda$ and are divergent in order
to absorb the one-loop ultraviolet divergences in dimensional regularization. Of these, only the LECs
$d_{16}$ and $d_{16}^i$ have finite pieces,
\begin{equation}
  d_{16}^{(i)} (\lambda) = d_{16}^{(i),r}(\lambda) + \frac{\beta^{(i)}_{16}}{F_\pi^2}L(\lambda)
  =\bar{d}_{16}^{(i)} + \frac{\beta^{(i)}_{16}}{F_\pi^2}\left(L(\lambda)+\frac{1}{(4\pi)^2}\ln\frac{M_\pi}{\lambda}\right) ,
\end{equation}
where $d_{16}^{(i),r}(\lambda)$ denote the renormalized, scale-dependent LECs, whereas $\bar{d}_{16}^{(i)}$ denote
the scale-independent counterparts. The terms $\propto \tilde{d}_{25,29}^{(i)}$ are only needed for the
absorption of divergences of the one-loop functional, so the corresponding LECs have no finite part,
 \begin{equation}
   \tilde{d}_{25,29}^{(i)} (\lambda) =\frac{\beta^{(i)}_{25,29}}{F_\pi^2}\left(L(\lambda)
   +\frac{1}{(4\pi)^2}\ln\frac{M_\pi}{\lambda}\right) .
\end{equation}
In these equations, $L(\lambda)$ contains the divergence at space-time dimension $d=4$,
\begin{equation}\label{eq:Ldivergence}
  L(\lambda)=\frac{\lambda^{d-4}}{(4\pi)^2}\left( \frac{1}{d-4}-\frac{1}{2}\left[\ln (4\pi)
    + \Gamma^\prime (1)+1\right]\right) ,
\end{equation}
and the $\beta$-functions are set to cancel the divergences of the one-loop functional, as discussed below.

In order to derive the full axion-nucleon coupling at $\mathcal{O}(p^3)$, the Lagrangian~\eqref{eq:LagrangianNPi}
has to be expressed in terms of the axion field $a$ and the matrix-valued field $u$ has to be expanded to the
required order. For the $\mathcal{O}(p^3)$ tree-level contribution, we hence can set $u=\mathbbm{1}$, whereas
for the $\mathcal{O}(p^3)$ pion-loop contributions, we have to expand $u$ to $\mathcal{O}(\pi^2)$.
Both calculations are done in the subsequent sections.

\subsection{Tree-level contributions at \texorpdfstring{$\mathcal{O}(p^3)$}{O(p3)}}

All interaction terms of the Lagrangian~\eqref{eq:LagrangianNPi} contribute. The expressions for
the external sources can be read off from the axion-quark interaction Lagrangian~\eqref{eq:a-q-lagrangian}:
\begin{align}
s & =  \mathcal{M}_a~, \nonumber\\
p & =  v_\mu = 0~, \\
a_\mu & =  c_{u-d} \frac{\partial_\mu a}{2f_a} \tau_3~,\nonumber\\
a_{\mu,i}^s & =   c_i \frac{\partial_\mu a}{2f_a} \mathbbm{1}\ .\nonumber
\end{align}
Setting hence $u=\mathbbm{1}$ and expanding the exponential in $\mathcal{M}_a$, cf. eq.~\eqref{eq:Mass:a},
up to $\mathcal{O}(a^1)$, we find
\begin{align}
D_\mu & =  \partial_\mu~,\nonumber\\
u_\mu & =  c_{u-d} \frac{\partial_\mu a}{f_a} \tau_3~,\nonumber\\
\tilde{u}_{\mu,i} & =  c_i \frac{\partial_\mu a}{f_a} \tau_3~,\\
\chi_{+} & =  4 B \mathcal{M}_q~,\nonumber\\
\chi_{-} & =  \frac{4iM_\pi^2}{f_a} \frac{m_u m_d}{(m_u+m_d)^2} a\ ,\nonumber
\end{align}
where we have inserted the leading order pion mass $M_\pi^2 = B(m_u+m_d)$.
Introducing the abbreviation
\begin{equation}
g_a = g_A c_{u-d} \tau_3 + g_0^i c_i \mathbbm{1} ,
\end{equation}
the single axion-nucleon interaction Lagrangian reads
\begin{eqnarray}
\mathcal{L}_{a N}^\text{int.} = \frac{1}{f_a} \bar{N} \Biggl\{ && 
		 g_a S\cdot (\partial a) - \frac{ig_a}{2m}\left\{S\cdot \partial, v\cdot (\partial a)\right\} \nonumber\\
		&& +\frac{g_a}{4m^2}\biggl(\overset{\leftarrow}{\partial}_\mu\,S\cdot (\partial a)\, \partial^\mu-v\cdot\overset{\leftarrow}{\partial}\, S\cdot(\partial a)\, v\cdot\partial \nonumber\\
		 && \qquad\qquad -\left(\left\{S\cdot\partial,v\cdot(\partial a)\right\}v\cdot\partial + \text{h.c.}\right)-\left(S\cdot\overset{\leftarrow}{\partial}\, (\partial a) \cdot \partial + \text{h.c.}\right)\biggr)\nonumber\\
           \qquad\qquad	& & + 4M_\pi^2\biggl(\left[d_{16}(\lambda)\tau_3+d_{17}\frac{m_u-m_d}{m_u+m_d}\right]c_{u-d} + d_{16}^i(\lambda) c_i 
           \label{eq:aNNinteractionLagr} \\
		&& \qquad\qquad-\left[ d_{18}+2d_{19}\right]\frac{m_u m_d}{(m_u+m_d)^2}\biggr) S\cdot (\partial a)\nonumber\\
		&& +\left(\tilde{d}_{25}(\lambda)c_{u-d}\tau_3+\tilde{d}_{25}^i(\lambda) c_i \right) v\cdot\overset{\leftarrow}{\partial}\, S\cdot (\partial a)\, v\cdot \partial \nonumber\\
		&& +\left(\tilde{d}_{29}(\lambda)c_{u-d}\tau_3+ \tilde{d}_{29}^i(\lambda) c_i \right) \left(S^\mu\left[v\cdot\partial , (\partial a)\right] v\cdot\partial + \text{h.c.} \right)\Biggr\} N\ . \nonumber
\end{eqnarray}
From that, we can derive the corresponding tree-level $NNa$-vertex Feynman rule,
\begin{align}
\raisebox{-0.8cm}{\includegraphics[height=1.8cm]{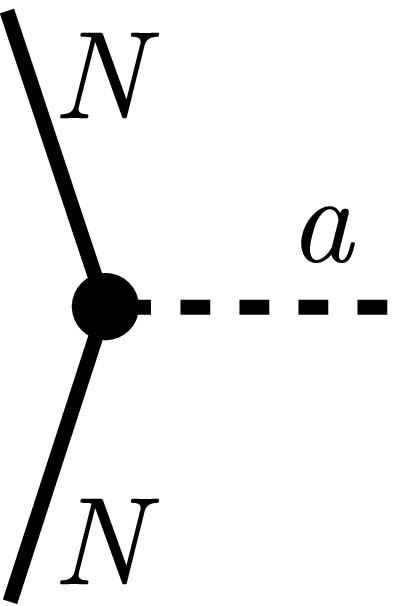}}\ = & -\frac{1}{f_a} \Biggl(
g_a \left[1+\frac{1}{2m}\left(\omega-\omega^\prime\right)-\frac{1}{4m^2}\left(\omega^2+\omega^\prime\left(\omega-\omega^\prime\right)-p^2\right)\right] \nonumber\\
& +4M_\pi^2 \biggl[\left(d_{16}(\lambda)\tau_3+d_{17}\frac{m_u-m_d}{m_u+m_d}\right) c_{u-d}
  + d_{16}^i(\lambda)c_i-\left(d_{18}+2d_{19}\right)\frac{m_um_d}{(m_u+m_d)^2}\biggr]\nonumber\\
& +\left[\tilde{d}_{25}(\lambda)c_{u-d}\tau_3
  +\tilde{d}_{25}^i(\lambda) c_i \right]\omega\omega^\prime + \left[\tilde{d}_{29}(\lambda)c_{u-d}\tau_3
  +\tilde{d}_{29}^i(\lambda) c_i\right]\left(\omega-\omega^\prime\right)^2 \Biggr) S\cdot q \nonumber \\
& +\frac{g_a}{mf_a} \left[(\omega-\omega^\prime)-\frac{1}{2m}(\omega^2-{\omega^\prime}^2)+\frac{1}{4m}(p^2-{p^\prime}^2)\right] S\cdot p \ , \label{eq:treelevelcontr}
\end{align}
where $q$ is the momentum of the outgoing axion, $p$ the momentum of the incoming nucleon, and $p^\prime = p-q$ the
momentum of the outgoing nucleon. Furthermore, we have set $\omega^{(\prime)}=v\cdot p^{(\prime)}$. 
Note that this expression~\eqref{eq:treelevelcontr} contains  divergences due to the terms $\propto
d_{16}^{(i)}(\lambda)$ and $\propto \tilde{d}_{25,29}^{(i)}(\lambda)$.

\subsection{Pion-loop contributions}

\begin{figure}[t]
\includegraphics[width=\textwidth]{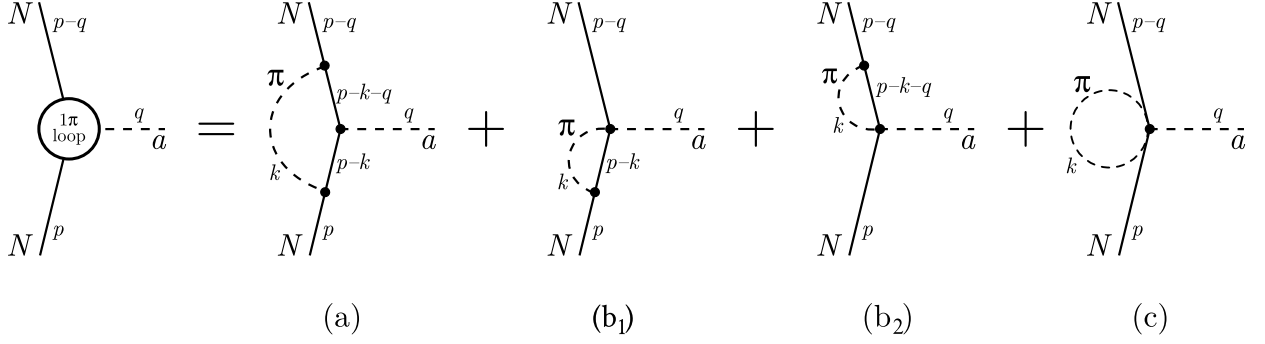}
\caption{Pion loop contributions to $N\to N+a$.}
\label{fig:oneloop}
\end{figure}

According to the usual power counting scheme, see, e.\,g., Ref.~\cite{Bernard:1995dp}, one-loop diagrams start
contributing at $\mathcal{O}(p^3)$. The relevant diagrams are shown in Fig.~\ref{fig:oneloop}. Note that axion loop
contributions are negligibly small due to the $1/f_a$ suppressions. We thus have to determine the $NN\pi$-,
$NN\pi a$-, and $NN\pi\pi a$-vertex Feynman rules from the leading order terms of the
Lagrangian~\eqref{eq:LagrangianNPi}. Expanding hence
\begin{equation}
  u=\exp\left(i\frac{\pi^a\tau_a}{2F_\pi}\right)= \mathbbm{1} + i\frac{\pi^a\tau_a}{2F_\pi}
  - \frac{\pi^a\tau_a \pi^b\tau_b}{8F_\pi^2} + \mathcal{O}(\pi^3)\ ,
\end{equation}
we find
\begin{align}
D_\mu & =  \partial_\mu + i\frac{c_{u-d}}{2f_aF_\pi} \partial_\mu a\, \epsilon_{3ab}\pi^a \tau^b~,\nonumber\\
u_\mu & =  - \frac{\partial_\mu \pi^a}{F_\pi} \tau_a + c_{u-d} \frac{\partial_\mu a}{f_a} \tau_3
+ c_{u-d} \frac{\partial_\mu a\, \pi^a\pi^b}{2f_aF_\pi^2} \left(\tau_a \delta_{3b} -\tau_3\delta_{ab} \right)~,\\
\tilde{u}_{\mu,i} & =  c_i \frac{\partial_\mu a}{f_a} \tau_3\ ,\nonumber
\end{align}
and thus
\begin{align}
  \mathcal{L}_{N\pi}^\text{int.} = \bar{N} \Biggl\{ & -\frac{1}{F_\pi} S\cdot (\partial \pi^a)\tau_a
  +\frac{1}{f_a}  (g_Ac_{u-d}\tau_3+g_0^ic_i) S\cdot (\partial a)  - \frac{c_{u-d}}{2f_aF_\pi} v\cdot
  (\partial a) \epsilon_{3ab} \pi^a\tau^b \nonumber\\
  &  +\frac{g_A c_{u-d}}{2f_a F_\pi^2} S\cdot (\partial a) \pi^a\pi^b \left(\tau_a \delta_{3b}
  -\tau_3\delta_{ab} \right) \Biggr\}N \ . 
\end{align}

\noindent
This yields the following Feynman rules with $k$ the outgoing pion momentum, and $q$ the outgoing axion momentum:
\begin{align*}
\text{pion propagator: } & \quad\frac{i\delta_{ab}}{k^2 -M_\pi^2 + i\eta}~, \\
\text{nucleon propagator: } & \quad\frac{i}{v\cdot p + i\eta}~, \\
NN\pi\text{-vertex: } & \quad\frac{g_A}{F_\pi} S\cdot k\, \tau_a~,\\
NNa\text{-vertex: } & \quad-\frac{1}{f_a} (g_A c_{u-d}\tau_3+g_0^i c_i) S\cdot q~,\\
NN\pi a\text{-vertex: } & \quad \frac{c_{u-d}}{2f_aF_\pi} v\cdot q\, \epsilon_{3ab}\tau_b~,\\
NN\pi\pi a\text{-vertex: } & \quad -\frac{g_Ac_{u-d}}{2f_aF_\pi^2} S\cdot q\,
\left(\tau_a \delta_{3b} -\tau_3\delta_{ab} \right) .
\end{align*}
In what follows, we will make use of the loop functions given in the appendix of Ref.~\cite{Bernard:1995dp}:
\begin{align}
  \Delta_\pi =-\frac{1}{i}\int\frac{\mathrm{d}^dk}{(2\pi)^d}\,\frac{1}{k^2-M_\pi^2+i\eta} & = 2M_\pi^2\left(L(\lambda)
  +\frac{1}{(4\pi)^2}\ln\frac{M_\pi}{\lambda}\right)+\mathcal{O}(d-4)~,\\
  \frac{1}{i}\int\frac{\mathrm{d}^dk}{(2\pi)^d}\,\frac{\left\{1,k_\mu,k_\mu k_\nu\right\}}{(k^2-M_\pi^2+i\eta)
    (\omega-v\cdot k+i\eta)} & = \left\{J_0(\omega),v_\mu J_1(\omega),g_{\mu\nu}J_2(\omega)+v_\mu v_\nu
  J_3(\omega)\right\}\ ,
\end{align}
where $L(\lambda)$ is given in eq.~\eqref{eq:Ldivergence}, and
\begin{align}
  J_0(\omega) & = -4\omega L + \frac{2\omega}{(4\pi)^2}\left(1-2\ln\frac{M_\pi}{\lambda}\right)
  -\frac{1}{4\pi^2}\sqrt{M_\pi^2-\omega^2}\arccos\frac{-\omega}{M_\pi}+\mathcal{O}(d-4)~, \\
J_1(\omega) & = wJ_0(\omega)+\Delta_\pi~,\\
J_2(\omega) & = \frac{1}{d-1} \left[(M_\pi^2-\omega^2)J_0(\omega)-\omega\Delta_\pi \right]~,\label{eq:loopfctJ2}\\
J_3(\omega) & = w J_1(\omega) - J_2(\omega)\ ,
\end{align}
where in particular the expression of $J_0(\omega)$ is valid for $\omega < M_\pi$, which is the region we are interested in.

\subsubsection{Diagram \texorpdfstring{(a)}{a}}

Using the Feynman rules given above, the first loop diagram is calculated as 
\begin{equation}\label{eq:diagrama}
  \text{(a)} = \frac{\hat{g}_a g_A^2}{f_aF_\pi^2}S^\mu\,S\cdot q\, S^\nu \frac{1}{i}\int
  \frac{\mathrm{d}^dk}{(2\pi)^d}\, \frac{k_\mu k_\nu}{(k^2-M_\pi^2+i\eta)}\frac{1}{(\omega^\prime
    - v\cdot k+i\eta)(\omega - v\cdot k+i\eta)}~,
\end{equation}
where we have set (summation over $i,j$ implied)
\begin{equation}
\hat{g}_a=\tau_i g_a \tau_j \delta_{ij} = - g_A c_{u-d} \tau_3 + 3 g_0^i c_i \mathbbm{1}\ .
\end{equation}
Using the identity
\begin{equation}
  \frac{1}{(\omega^\prime - v\cdot k+i\eta)(\omega - v\cdot k+i\eta)} = \frac{1}{\omega^\prime-\omega}
  \left[\frac{1}{\omega - v\cdot k+i\eta}-\frac{1}{\omega^\prime - v\cdot k+i\eta} \right],
\end{equation}
the equation~\eqref{eq:diagrama} can be written as
\begin{align}
  \text{(a)} & =  \frac{\hat{g}_a g_A^2}{f_aF_\pi^2}S^\mu\,S\cdot q\, S^\nu\frac{1}{\omega^\prime-\omega}
  \left(\frac{1}{i} \int \frac{\mathrm{d}^dk}{(2\pi)^d}\, \frac{k_\mu k_\nu}{(k^2-M_\pi^2+i\eta)(\omega
    - v\cdot k+i\eta)} - \omega\to\omega^\prime\right)\nonumber\\
  & = \frac{\hat{g}_a g_A^2}{f_aF_\pi^2} S^\mu\,S\cdot q\, S^\nu \frac{1}{\omega^\prime-\omega}
  \left(g_{\mu\nu} \left[J_2(\omega)-J_2(\omega^\prime)\right]+  v_{\mu} v_{\nu} \left[J_3(\omega)
    -J_3(\omega^\prime)\right]\right) .
\end{align}
The terms $\propto J_3(\omega^{(\prime)})$ vanish because of Eq.~\eqref{eq:Spinproperties}. The final result
is found using the anticommutator~\eqref{eq:Spinproperties} and inserting the loop function \eqref{eq:loopfctJ2},
\begin{align}
  \text{(a)} =  & \frac{\hat{g}_a}{6f_a}\left(\frac{g_A}{4\pi F_\pi}\right)^2 \Biggl\{-M_\pi^2
  +\frac{1}{\omega-\omega^\prime}\biggl(\omega^3-{\omega^\prime}^3 \nonumber\\ & \qquad\qquad
  +2\left[\left(M_\pi^2-\omega^2\right)^\frac{3}{2}\arccos\frac{-\omega}{M_\pi}-\left(M_\pi^2
    -{\omega^\prime}^2\right)^\frac{3}{2}\arccos\frac{-\omega^\prime}{M_\pi}\right]\biggr)\Biggr\} S\cdot q \nonumber\\
  & + \frac{\hat{g}_a g_A^2}{6f_aF_\pi^2}\left(3M_\pi^2-2\left(\omega-\omega^\prime\right)^2
  -6\omega\omega^\prime \right)\left(L(\lambda)+\frac{1}{(4\pi)^2}\ln \frac{M_\pi}{\lambda}\right)S\cdot q\ , 
\end{align}
where we have separated the finite, scale-independent terms (the first and second lines) from the divergent or
scale-dependent ones (the third line).

\subsubsection{Diagrams \texorpdfstring{(b\textsubscript{1}) and (b\textsubscript{2})}{b1 and b2}}

Diagrams (b\textsubscript{1}) and (b\textsubscript{2}) have the same structure:
\begin{equation}
  \text{(b\textsubscript{1})}\propto S^\mu \frac{1}{i} \int\frac{\mathrm{d}^dk}{(2\pi)^d}\,
  \frac{k_\mu}{(M_\pi^2-k^2+i\eta)(\omega-v\cdot k +i\eta)}= S\cdot v\, J_1(\omega)=0 \ ,
\end{equation}
where we once again made use of $S\cdot v=0$, see~\eqref{eq:Spinproperties}. For diagram (b\textsubscript{2})
one just needs to replace $\omega\to\omega^\prime$.

\subsubsection{Diagram \texorpdfstring{(c)}{c}}
The last diagram is divergent:
\begin{equation}
  \text{(c)} = -\frac{g_Ac_{u-d}}{f_aF_\pi^2} S\cdot q\, \frac{1}{i}\int\frac{\mathrm{d}^dk}{(2\pi)^d}\,
  \frac{1}{k^2-M_\pi^2+i\eta}  = \frac{2g_Ac_{u-d}M_\pi^2}{f_aF_\pi^2}\left(L(\lambda)+\frac{1}{(4\pi)^2}
  \ln\frac{M_\pi}{\lambda}\right)\, .
\end{equation}

\subsection{Axion-nucleon coupling at \texorpdfstring{$\mathcal{O}(p^3)$}{Op3}}
In order to remove the  divergences appearing in the coupling~\eqref{eq:treelevelcontr} and diagrams (a) and (c),
we utilize the following set of $\beta$-functions:
\begin{equation}
\begin{array}{rlrlrl}
\beta_{16} & = \frac{g_A}{8}(4-g_A^2)\ , & \beta_{25} & = g_A^3\ , &
\beta_{29} & = \frac{g_A^3}{3}\ , \\
\beta_{16}^i & = \frac{3}{8}g_A^2g_0^i\ , & \beta_{25}^i & = -3g_A^2g_0^i\ , & \beta_{29}^i & = -g_A^2g_0^i\ ,
\end{array}
\end{equation}
where $\beta_{16}$, $\beta_{25}$, and $\beta_{29}$ have been calculated already before within the theory
without axions~\cite{Ecker:1994pi}. The remaining $\beta$-functions $\beta_{16}^i$, $\beta_{25}^i$, and
$\beta_{29}^i$ are new in the theory with axions and have been worked out  here for the first time.
We thus have a finite $NNa$-vertex with the Feynman rule:
\begin{align}
\raisebox{-0.8cm}{\includegraphics[height=1.8cm]{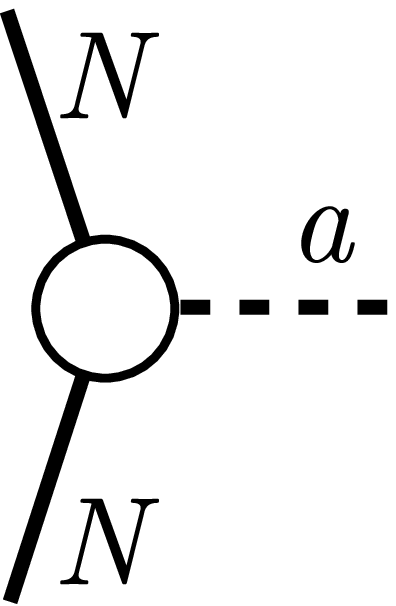}}\ = & -\frac{1}{f_a} \Biggl(
g_a \left[1+\frac{1}{2m}\left(\omega-\omega^\prime\right)-\frac{1}{4m^2}\left(\omega^2+\omega^\prime\left(\omega-\omega^\prime\right)-p^2\right)\right]-\frac{\hat{g}_a}{6}\left(\frac{g_A }{4\pi F_\pi}\right)^2 \nonumber\\  & \quad\times \biggl[M_\pi^2 -\frac{1}{\omega-\omega^\prime} \biggl(\omega^3-{\omega^\prime}^3 +2\left[\left(M_\pi^2-\omega^2\right)^\frac{3}{2}\arccos\frac{-\omega}{M_\pi}-\left(M_\pi^2-{\omega^\prime}^2\right)^\frac{3}{2}\arccos\frac{-\omega^\prime}{M_\pi}\right]\biggr)\biggr] \nonumber\\ & + 4M_\pi^2 \biggl[\left(\bar{d}_{16} \tau_3+d_{17}\frac{m_u-m_d}{m_u+m_d}\right) c_{u-d} + \bar{d}_{16}^i c_i-\left(d_{18}+2d_{19}\right)\frac{m_um_d}{(m_u+m_d)^2}\biggr] \Biggr) S\cdot q \nonumber \\
		& +\frac{g_a}{mf_a} \left[(\omega-\omega^\prime)-\frac{1}{2m}(\omega^2-{\omega^\prime}^2)+\frac{1}{4m}(p^2-{p^\prime}^2)\right] S\cdot p \ , \label{eq:orderp3axionnucleonvertex}
\end{align}
For the following estimation of the coupling strengths of the axion-proton and the axion-neutron vertex,
we assume the rest frame of the incoming nucleon, i.e. $v=(1,0,0,0)^\text{T}$, so that $\omega=p=0$ and
$\omega^\prime=-v \cdot q \ll m$. In the rest frame, Eq.~\eqref{eq:orderp3axionnucleonvertex} becomes
\begin{align}
\raisebox{-0.8cm}{\includegraphics[height=1.8cm]{NNa_vertex}}\ & =  -\frac{1}{f_a} \Biggl(
g_a \left[1-\frac{\omega^\prime}{2m}+\left(\frac{\omega^\prime}{2m}\right)^2\right]\nonumber\\  
& \qquad + \frac{\hat{g}_a}{6}\left(\frac{g_A M_\pi}{4\pi F_\pi}\right)^2 \left[-1 +\left(\frac{\omega^\prime}{M_\pi}\right)^2 - \frac{2}{\omega^\prime M_\pi^2}\left(\frac{\pi M_\pi^3}{2} -\left(M_\pi^2-{\omega^\prime}^2\right)^\frac{3}{2}\arccos\frac{-\omega^\prime}{M_\pi}\right)\right] \nonumber\\ 
& \qquad + 4M_\pi^2 \biggl[\left(\bar{d}_{16} \tau_3+d_{17}\frac{m_u-m_d}{m_u+m_d}\right) c_{u-d} + \bar{d}_{16}^i c_i-\left(d_{18}+2d_{19}\right)\frac{m_um_d}{(m_u+m_d)^2}\biggr] \Biggr) S\cdot q  \nonumber \\
& = -\frac{1}{f_a} \left(g_a f_1(w^\prime)+ \frac{\hat{g}_a}{6}\left(\frac{g_A M_\pi}{4\pi F_\pi}\right)^2 f_2(w^\prime) + g_a^\text{N\textsuperscript{2}LO} \right) S\cdot q \ ,
 \label{eq:orderp3axionnucleonvertexinrestframe}
\end{align}
where 
\begin{align}
f_1(\omega^\prime) & = 1-\frac{\omega^\prime}{2m}+\left(\frac{\omega^\prime}{2m}\right)^2 ,\\
f_2(\omega^\prime) & = -1 +\left(\frac{\omega^\prime}{M_\pi}\right)^2 - \frac{2}{\omega^\prime M_\pi^2}\left(\frac{\pi M_\pi^3}{2} -\left(M_\pi^2-{\omega^\prime}^2\right)^\frac{3}{2}\arccos\frac{-\omega^\prime}{M_\pi}\right),\\
g_a^\text{N\textsuperscript{2}LO} & = 4 M_\pi^2 \biggl[\left(\bar{d}_{16} \tau_3+d_{17}\frac{m_u-m_d}{m_u+m_d}\right) c_{u-d} + \bar{d}_{16}^i c_i-\left(d_{18}+2d_{19}\right)\frac{m_um_d}{(m_u+m_d)^2}\biggr] \ .
\end{align}
Since we are working in the very-low-energy regime, the function $f_2(w^\prime)$ may be approximated for $\omega^\prime \ll  M_\pi$. A series expansion around $\omega^\prime = 0$ yields
\begin{equation}
f_2(\omega^\prime) = 1-\frac{3\pi}{2}\frac{\omega^\prime}{M_\pi} - \frac{5}{3}\left(\frac{\omega^\prime}{M_\pi}\right)^2 + \mathcal{O}\left(\left(\frac{\omega^\prime}{M_\pi}\right)^3\right) ,
\end{equation}
so that we find the axion-nucleon coupling at zero momentum transfer
\begin{equation}\label{eq:axionnucleoncouplingeneral}
G_{aNN} = -\frac{1}{f_a} g_{aNN} = -\frac{1}{f_a} \left(g_a + g_a^\text{loop} + g_a^\text{N\textsuperscript{2}LO} \right),
\end{equation}
with
\begin{equation}
g_{a}^\text{loop} = \frac{\hat{g}_a}{6}\left(\frac{g_A M_\pi}{4\pi F_\pi}\right)^2  .
\end{equation}
The respective axion-proton and axion-neutron vertices can be determined by inserting the
expressions for $c_{u-d}$ and the $c_i$'s from Eq.~\eqref{eq:X3andXs} and Eq.~\eqref{eq:cqs} and
by matching $g_A$ and the $g_0^i$'s to the nucleon matrix elements, i.\,e.
\begin{eqnarray}
g_A &=& \Delta u-\Delta d~, \nonumber \\
g_0^{u+d} &=& \Delta u+\Delta d~, \\
g_0^q &=& \Delta q~,\text{ for } q=s,c,b,t\ , \nonumber
\end{eqnarray}
where $s^\mu\Delta q=\langle p | \bar{q}\gamma^\mu\gamma_5 q|p\rangle$, with $s^\mu$ the spin of the proton.
The proton and neutron matrix elements are related by isospin symmetry, i.e. $\langle p | \bar{u}\gamma^\mu\gamma_5
u|p\rangle = \langle n | \bar{d}\gamma^\mu\gamma_5 d|n\rangle$, $\langle p | \bar{d}\gamma^\mu\gamma_5 d|p\rangle
=\langle n | \bar{u}\gamma^\mu\gamma_5 u|n\rangle$, and $\langle p | \bar{q}\gamma^\mu\gamma_5 q|p\rangle=\langle
n | \bar{q}\gamma^\mu\gamma_5 q|n\rangle$ for $q=s,c,b,t$. In particular we find
\begin{align}
g_a^p & = -\frac{\Delta u+z\Delta d+w\Delta s}{1+z+w}+\Delta u X_u + \Delta d X_d+\sum_{q=\{s,c,b,t\}} \Delta q X_q ~,\\
\hat{g}_a^p & = -\frac{(1+2z)\Delta u+(2+z)\Delta d+3w\Delta s}{1+z+w} \nonumber\\
& \qquad\qquad +(\Delta u+2\Delta d) X_u + (2\Delta u+\Delta d) X_d+3\sum_{q=\{s,c,b,t\}} \Delta q X_q ~,
\end{align}
for the case of axion-proton interaction, and
\begin{align}
g_a^n & = -\frac{z\Delta u+\Delta d+w\Delta s}{1+z+w}+\Delta d X_u + \Delta u X_d+\sum_{q=\{s,c,b,t\}} \Delta q X_q ~,\\
\hat{g}_a^n & = -\frac{(2+z)\Delta u+(1+2z)\Delta d+3w\Delta s}{1+z+w}\nonumber\\ & \qquad\qquad +(2\Delta u+\Delta d) X_u
+ (\Delta u+2\Delta d) X_d+3\sum_{q=\{s,c,b,t\}} \Delta q X_q ~,
\end{align}
for the case of axion-neutron coupling.

We extract the respective quantities from the recent FLAG review \cite{Aoki:2019cca}, which are here given in the
$\overline{\text{MS}}$ scheme at the scale $\mu=2\,\text{GeV}$ (we utilize the  nucleon matrix elements
calculated on the lattice with $N_f=2+1$ excluding isospin breaking effects). The LEC $d_{18}$ is
taken from Ref.~\cite{Hoferichter:2015tha} (fixed by the Goldberger-Treiman discrepancy), whereas $\bar{d}_{16}$ has been adopted from Ref.~\cite{Siemens:2017opr}:
\begin{equation}
\begin{array}{rlrlrl}
\Delta u & = 0.847(50) , & \Delta d & = -0.407(34) , & \Delta s & = -0.035(13)  , \\
m_u & = 2.27(9)\,\text{MeV} , & m_d & = 4.67(9)\,\text{MeV}  , & M_\pi & =136.10(1.82)\,\text{MeV} , \\
z & = 0.485(19) , & w & = 0.025(1)  , &  \mathring{m}_N & =0.8726(31)\,\text{GeV}, \\
\bar{d}_{16} & = 0.4(1.3)\,\text{GeV}^{-2}, & d_{18} & =-0.44(24)\,\text{GeV}^{-2}\ , & & \\
\end{array}
\end{equation}
where the nucleon mass in the chiral limit $m=\mathring{m}_N$ has been estimated via the
third order relation (which is the accuracy to which we are working)
\begin{equation}
  m_{N,\text{phys}} = \mathring{m}_N - 4c_1M_\pi^2 - \frac{3g_A^2M_\pi^3}{32\pi F_\pi^2} + {\cal O}(M_\pi^4)~,
\end{equation}
with $c_1 = -1.07(2)$~GeV$^{-1}$ at ${\cal O}(p^3)$ from Ref.~\cite{Hoferichter:2015tha}.
Note that the pion mass is taken to be the leading order pion mass $M_\pi=\sqrt{B(m_u+m_d)}$.
The $\bar{d}_{16}^i$'s are of course hitherto undetermined since they are new LECs in the theory with axions.

Inserting these values, one finds
\begin{align}
g_a^p & = -0.430(36)+0.847(50)X_u-0.407(34)X_d-0.035(13)X_s~, \label{eq:LOpcoupling}\\
\hat{g}_a^p & = -0.433(36)+0.033(84)X_u+1.287(106)X_d-0.105(39)X_s~,\\
g_a^{p,\text{loop}} & = -0.002(1)+0.0001(3)X_u+0.005(2)X_d-0.0004(2)X_s~,
\end{align}
for the case of axion-proton interaction, and
\begin{align}
g_a^n & = -0.002(30)-0.407(34)X_u+0.847(50)X_d-0.035(13)X_s~, \label{eq:LOncoupling} \\
\hat{g}_a^n & =  -0.861(30)+1.287(106)X_u+0.033(84)X_d-0.105(39)X_s~,\\
g_a^{n,\text{loop}} & =  -0.003(1)+0.005(2)X_u+0.0001(3)X_d-0.0004(2)X_s~,
\end{align}
for the interaction of axions with neutrons.
Note that $g_a^p$ and $g_a^n$ are nothing but the pure leading order coupling strengths, which were reported
already in Ref.~\cite{diCortona:2015ldu} (Eq.~(2.49) in their paper). Here we have used the most recent values
for the involved quantities and neglected terms $\propto \Delta c,\Delta b,\Delta t$, because these contributions
are well beyond the accuracy of the present N$^2$LO estimations of the coupling strengths.

The corrections to the leading order couplings~\eqref{eq:LOpcoupling} and \eqref{eq:LOncoupling} stemming from
the chiral expansion to N$^2$LO is given by
\begin{align}
  g_a^{p, \text{N\textsuperscript{2}LO}} & = 0.002(7) \nonumber \\ & \quad +\left(-0.036(1)\bar{d}_{16}^{u+d}-0.001(0)\bar{d}_{16}^s
  +0.004(0)d_{17}+0.033(1)d_{19}\right)\text{GeV}^2 \nonumber\\
& \quad + \left( 0.015(48)+\left[0.037(1)\bar{d}_{16}^{u+d}-0.013(1)d_{17}\right]\text{GeV}^2\right) X_u \nonumber\\
& \quad + \left( - 0.015(48)+\left[0.037(1)\bar{d}_{16}^{u+d}+0.013(1)d_{17}\right]\text{GeV}^2\right) X_d \nonumber\\
& \quad +0.074(2)\text{GeV}^2 \bar{d}_{16}^s X_s\ ,\\
  g_a^{n, \text{N\textsuperscript{2}LO}} & = 0.012(7) \nonumber\\ & \quad +\left(-0.036(1)\bar{d}_{16}^{u+d}-0.001(0)\bar{d}_{16}^s
  +0.004(0)d_{17}+0.033(1)d_{19}\right)\text{GeV}^2 \nonumber\\
&  \quad+ \left(- 0.015(48)+\left[0.037(1)\bar{d}_{16}^{u+d}-0.013(1)d_{17}\right]\text{GeV}^2\right) X_u \nonumber\\
&  \quad+ \left( 0.015(48)+\left[0.037(1)\bar{d}_{16}^{u+d}+0.013(1)d_{17}\right]\text{GeV}^2\right) X_d\nonumber \\
&  \quad+0.074(2)\text{GeV}^2 \bar{d}_{16}^s X_s\ .
\end{align}
Since the
values of the remaining LECs appearing in this expression are unknown, we  estimate the strength of the contribution
of $g_a^\text{N\textsuperscript{2}LO}$ by assuming that the undetermined LECs are of $\mathcal{O}(\text{GeV}^{-2})$
following the conventional naturalness arguments, see e.g.~\cite{Siemens:2017opr, Bernard:2007zu}. We expect the values
of these LECs to be comparable to the ones known from $\bar{d}_{16}$ and $d_{18}$, so we make the \emph{ansatz}
$|\bar{d}_{16}^i|=0.5(5)\text{GeV}^{-2}$ and likewise $|d_{17,19}|=0.5(5)\text{GeV}^{-2}$, and perform a Monte Carlo
simulation assuming a normal distribution for each undetermined LEC. This yields
\begin{align}
g_a^{p, \text{N\textsuperscript{2}LO}} & = 0.002(35) + 0.015(56) X_u - 0.015(56) X_d  + 0.000(52) X_s\ ,\\
g_a^{n, \text{N\textsuperscript{2}LO}} & = 0.012(35) - 0.015(56) X_u + 0.015(56) X_d  + 0.000(52) X_s\ ,
\end{align}
so that our final result for the axion-nucleon coupling \eqref{eq:axionnucleoncouplingeneral} reads
\begin{align}
g_{app} & = -0.430(50) + 0.862(75) X_u - 0.417(66) X_d  - 0.035(54) X_s\ ,\\
g_{ann} & = 0.007(46) - 0.417(66) X_u + 0.862(75) X_d  - 0.035(54) X_s\ .
\end{align}
Note that we are still working at the  matching scale $\mu=2\,\text{GeV}$ (in contrast to \cite{diCortona:2015ldu}).

Collecting all contributions, one gets for the KSVZ model with $X_q=0$,
\begin{align}
g_{app}^\text{KSVZ} &  = -0.430(50) \label{eq:gappKSVZ}~,\\
g_{ann}^\text{KSVZ} &  =  0.007(46) \label{eq:gannKSVZ}~,
\end{align}
while for the DFSZ axion,
\begin{align}
  g_{app}^\text{DFSZ}  & =  - 0.581(58) + 0.438(38) \sin^2\beta  ~,
  \label{eq:gappDSFZ}\\
  g_{ann}^\text{DFSZ} & =  0.283(55) - 0.415(38)\sin^2\beta ~.
  \label{eq:gannDSFZ}
\end{align}
The coupling of axions to nucleons hence is always non-zero in both models, even though $g_{ann}^\text{KSVZ}~=~0$ is
possibile within the error range. In the DFSZ model, the strength of the coupling to protons 
$g_{app}$ can range from $- 0.581(58)$ at $\sin^2\beta=0$ to $- 0.143(69)$
at $\sin^2\beta=1$. The coupling to neutrons may take on values from $+ 0.283(55)$ at
$\sin^2\beta=0$ to $- 0.132(67)$ at $\sin^2\beta=1$, which means that in the DSFZ model $g_{ann}$
might still vanish depending on the value of $\beta$.

\section{Summary}
\label{sec:summ}
In this work, we have calculated the axion-nucleon couplings at the next-to-next-to-leading order
in two-flavor non-relativistic baryon chiral perturbation theory. Including all phenomenological knowledge
from a variety of sources, we find the N$^2$LO corrections of a few percent only. These couplings are therefore
pinned down to a high precision.

Although we have reached a higher accuracy in the framework of chiral perturbation theory, the errors are still
relatively large for two reasons. Firstly, there are still sizeable uncertainties stemming from the LO nucleon matrix
elements calculated on the lattice, and secondly there are considerable uncertainties from the undetermined LECs.
In fact, a possible future detection of the axion could be used to determine these LECs by applying the method of Bayesian inference. This is, however, not
expected to be the case in the near future, because one would need more precise determinations of all other
involved quantities such as the nucleon matrix elements $\Delta q$ or the quark masses (which might be achieved
in a few years in lattice QCD), and at the same time very precise measurements of the axion-nucleon coupling would
be necessary. The formula in Eq.~\eqref{eq:orderp3axionnucleonvertex} is hence primarily of relevance particularly
for any future study on the axion-nucleon interaction, since the numerical values for the respective couplings can
always be brought up to date by using the presented formulas and inserting the most recent estimations for the
involved quantities.

In the future, it would be interesting to work out explicitly the chiral  corrections to the axion-photon
interaction and the influence of the strange quark.

\begin{acknowledgments}
This work is supported in part by the National Natural Science Foundation of China (NSFC) and  the
Deutsche Forschungsgemeinschaft (DFG) through the funds provided to the Sino-German Collaborative
Research Center ``Symmetries and the Emergence of Structure in QCD"  (NSFC Grant No. 11621131001,
DFG Grant No. TRR110), by the NSFC under Grant No. 11835015 and No. 11947302, by the Chinese Academy of
Sciences (CAS) under Grant No. QYZDB-SSW-SYS013 and No. XDPB09, by the CAS Center for Excellence in
Particle Physics (CCEPP),  by the CAS President's International Fellowship Initiative (PIFI) (Grant No.~2018DM0034),
and by the VolkswagenStiftung (Grant No. 93562).
\end{acknowledgments}

\end{document}